\documentclass[10pt,aps,prl,twocolumn,preprintnumbers,amsmath,amssymb,floatfix,showpacs,citeautoscript,superscriptaddress]{revtex4-2} %
\usepackage[latin1]{inputenc}
\usepackage[T1]{fontenc}
\usepackage{ulem}
\usepackage[english]{babel}
\usepackage[pdftex]{graphicx}
\usepackage{amsmath}
\usepackage{times}
\usepackage[usenames,dvipsnames,svgnames,table]{xcolor}
\usepackage[colorlinks,plainpages=false,linkcolor=blue,urlcolor=blue,citecolor=blue,pdfpagemode=UseNone]{hyperref}
\usepackage{hyperref}

\renewcommand{\AA}{\text{\r{A}}}

\definecolor{magenta}{rgb}{0.9,0.0,0.9}

\begin{document}

\title
{
\boldmath
Fermi surface reconstruction and enhanced spin fluctuations in strained La$_3$Ni$_2$O$_{7}$ \\ on LaAlO$_3$(001) and SrTiO$_3$(001)
}

\author{Benjamin Geisler}
\email{benjamin.geisler@ufl.edu}
\affiliation{Department of Physics, University of Florida, Gainesville, Florida 32611, USA}
\affiliation{Department of Materials Science and Engineering, University of Florida, Gainesville, Florida 32611, USA}
\author{James J. Hamlin}
\affiliation{Department of Physics, University of Florida, Gainesville, Florida 32611, USA}
\author{Gregory R. Stewart}
\affiliation{Department of Physics, University of Florida, Gainesville, Florida 32611, USA}
\author{Richard G. Hennig}
\affiliation{Department of Materials Science and Engineering, University of Florida, Gainesville, Florida 32611, USA}
\affiliation{Quantum Theory Project, University of Florida, Gainesville, Florida 32611, USA}
\author{P.J. Hirschfeld}
\affiliation{Department of Physics, University of Florida, Gainesville, Florida 32611, USA}

\date{\today}

\begin{abstract}
We explore the structural and electronic properties of the bilayer nickelate La$_3$Ni$_2$O$_{7}$ on LaAlO$_3$(001) and SrTiO$_3$(001)
by using density functional theory including a Coulomb repulsion term.
For La$_3$Ni$_2$O$_{7}$/LaAlO$_3$(001), we find that compressive strain and electron doping across the interface
result in the unconventional occupation of the antibonding Ni $3d_{z^2}$ states.
In sharp contrast, no charge transfer is observed for La$_3$Ni$_2$O$_{7}$/SrTiO$_3$(001).
Surprisingly, tensile strain drives a metallization of the bonding Ni $3d_{z^2}$ states, %
rendering a Fermi surface topology akin to superconducting bulk La$_3$Ni$_2$O$_{7}$ under high pressure,
yet with spin fluctuations enhanced considerably beyond pressure effects.
Concomitantly, significant octahedral rotations are retained.
We discuss the fundamental differences between hydrostatic pressure versus epitaxial strain
and establish that strain provides a much stronger control over the Ni~$e_g$ orbital polarization.
The results suggest epitaxial La$_3$Ni$_2$O$_{7}$, particularly under tensile strain, as interesting system
to provide novel insights into the physics of bilayer nickelates
and possibly induce superconductivity without external pressure.
\end{abstract}

\maketitle

\textit{Introduction. --}
The recent discovery of superconductivity with $T_c \sim 80$~K in pressurized La$_3$Ni$_2$O$_7$ \cite{Sun-327-Nickelate-SC:23, Hou-LNO327-ExpConfirm:23, Zhang-LNO327-ZeroResistance:23}
positioned the bilayer Ruddlesden-Popper compounds as an exciting new addition to the expanding class of superconducting nickelates,
drawing significant attention~\cite{Luo-LNO327:23, Gu-LNO327:23, Yang-LNO327:23, Lechermann-LNO327:23, Sakakibara-LNO327:23, Shen-LNO327:23, Christiansson-LNO327:23, Shilenko-LNO327:23, Wu-LNO327:23, Cao-LNO327:23, Chen-LNO327:23, Lu-LNO327-InterlayerAFM:23, ZhangDagotto-LNO327:23, Liao-LNO327:23, Qu-LNO327:23, HuangZhou-LNO327:23, QinYang-LNO327:23, Liu-LNO327-OxVacDestructive:23, ZhangDagotto-RE-LNO327:23, Liu-LNO327-Optics:23, Geisler-LNO327-Structure:23, RhodesWahl-LNO327:23, Wang-LNO327-SC-OxDeficient:23, Yang-LNO327-ARPES:23, Lu-LNO327:23, Wang-LNO327-SC:23, Wang-LNO327-I4mmm:23, Chen-LNO327-SDW:23, ZhengWu-LNO327:23, Kakoi-LNO327:23, Geisler-LNO327-Optical:24, LNO-327-Dong-VO-Visualization:24, Chen-Mitchell-Stacking-LNO327:24, Wang-La2PrNi2O7:24, LNO-327-SDW-LaBollita:24, Huo-LNO-Strain:25}.
La$_3$Ni$_2$O$_7$ is particularly intriguing as it exhibits superconductivity in bulk crystals, %
in contrast to infinite-layer nickelates~\cite{Li-Supercond-Inf-NNO-STO:19, Li-NoSCinBulkDopedNNO:19, Botana-Inf-Nickelates:19, Li-Supercond-Dome-Inf-NNO-STO:20, Zeng-Inf-NNO:20, Osada-PrNiO2-SC:20, Wang-NoSCinBulkDopedNNO:20, Osada-LaNiO2-SC:21, GoodgeGeisler-NNO-IF:22}.
The pairing mechanism has been suggested to be related
to a pressure-driven change in Fermi surface topology~\cite{Sun-327-Nickelate-SC:23, Luo-LNO327:23, Gu-LNO327:23, Yang-LNO327:23, ZhangDagotto-LNO327:23}
and a concomitant %
orthorhombic-to-tetragonal phase transition~\cite{Geisler-LNO327-Structure:23, Wang-LNO327-I4mmm:23}
involving a suppression of the octahedral rotations~\cite{Sun-327-Nickelate-SC:23, Geisler-LNO327-Structure:23, ZhangDagotto-LNO327:23, Wang-LNO327-I4mmm:23}.
Despite strong efforts, sample growth and characterization under high pressure %
remain challenging,
particularly in verifying the Meissner effect and detecting thermodynamic transitions. %

It is therefore essential to assess alternative pathways to drive superconductivity in bilayer nickelates. %
Here we investigate the impact of epitaxial strain %
as well as electrostatic doping exploiting interface polarity.
To this end, we explore the structural and electronic properties of La$_3$Ni$_2$O$_{7}$ on
LaAlO$_3$(001) (LAO) and SrTiO$_3$(001) (STO) %
by using density functional theory including a Coulomb repulsion term,
treating the interface explicitly.
We identify the fundamental differences between hydrostatic pressure and biaxial strain, %
and show that they probe orthogonal domains in crystal geometry and Ni~$3d$ orbital occupation. %
For La$_3$Ni$_2$O$_{7}$/LAO(001), compressive strain %
results in the unconventional occupation of the antibonding Ni~$3d_{z^2}$ states,
which we find to be further promoted by electron doping across the interface.
In sharp contrast, no charge transfer is observed for La$_3$Ni$_2$O$_{7}$/STO(001).
The Fermi energy is located in the substrate band gap in both systems,
such that the active states arise exclusively from the Ni~$e_g$ orbitals.
Counter-intuitively, we find that tensile strain in La$_3$Ni$_2$O$_7$/STO(001) %
renders a Fermi surface topology comparable to bulk La$_3$Ni$_2$O$_{7}$ under pressure, %
characterized by the emergence of a Ni~$3d_{z^2}$ hole pocket.
Simultaneously, the octahedral rotations are preserved.
The concomitant enhancement of the dynamical spin susceptibility exceeds even the strong effects of pressure and suggests particularly
bilayer nickelates under tensile strain %
as interesting platform to gain deeper insights into the pairing mechanism
and as potential candidate for superconductivity at ambient pressure. %

\begin{figure*}
\begin{center}
\includegraphics[width=\linewidth]{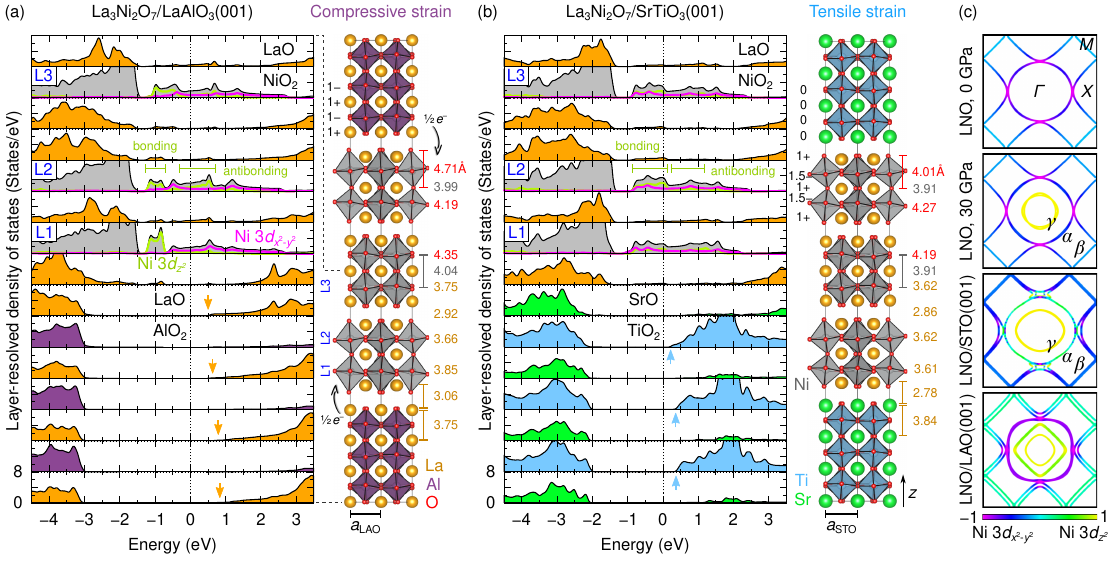}
\caption{\label{fig:OptGeo}
\textbf{\boldmath Optimized geometry, electronic structure, and Fermi surfaces of the bilayer nickelate systems considered here.}
(a,b)~Optimized geometry and layer-resolved density of states of La$_3$Ni$_2$O$_7$ (LNO) on
LAO(001) (compressive strain) and STO(001) (tensile strain).
The numbers on the right of each structure denote the $A$- (orange) and Ni-site distances (grey) as well as the O-Ni-O bond lengths (red) in $z$~direction.
The formal layer polarity is given by the black numbers.
Both substrates show no involvement in the Fermi surface,
although a downwards bending of their conduction-band states near the interface can be observed (orange and blue arrows).
(c)~Fermi surfaces, colored by the Ni orbital character $(3d_{z^2}-3d_{x^2-y^2})/(3d_{z^2}+3d_{x^2-y^2})$.
The orbital character of the $\alpha$ and $\beta$ sheets is considerably modulated by strain, exceeding the effect of external pressure.
On LAO, the central pockets stem from the antibonding Ni~$3d_{z^2}$ states
due to a concerted effect of compressive strain and charge transfer across the interface.
In sharp contrast, the STO substrate drives a metallization of the bonding Ni~$3d_{z^2}$ states ($\gamma$), resembling pressurized bulk La$_3$Ni$_2$O$_7$.
}
\end{center}
\end{figure*}

\textit{Methods. --}
We performed density functional theory calculations~\cite{KoSh65} (DFT$+U$)
by using Quantum ESPRESSO~\cite{PWSCF, QE-LDA-U:05, Vanderbilt:1990},
employing the PBE exchange-correlation functional~\cite{PeBu96} %
and $U=4$~eV on Ni and Ti sites to describe static correlation effects~\cite{GeislerPentcheva-InfNNO:20, GeislerPentcheva-NNOCCOSTO:21, Geisler-VO-LNOLAO:22, Sun-327-Nickelate-SC:23, ZhangDagotto-LNO327:23}.
To account for octahedral rotations,
we use $\sqrt{2}a \times \sqrt{2}a \times c$ supercells with two transition metal sites per layer,
setting $a$ to the substrate lattice parameter ($a_\text{LAO} = 3.79~\AA$, $a_\text{STO} = 3.905~\AA$)
and accurately optimizing $c$ as well as the ionic positions.
The symmetric supercells contain two equivalent interfaces, 3 bilayers of La$_3$Ni$_2$O$_7$ (LNO), and 6.5 layers of substrate,
corresponding to 136 atoms (Fig.~\ref{fig:OptGeo}). %
Reference calculations for bulk LNO at 0~GPa ($Cmcm$) and 30~GPa ($I4/mmm$) were performed
by using 48-atom supercells~\cite{Geisler-LNO327-Structure:23, Geisler-LNO327-Optical:24}.
Tight-binding Hamiltonians describing the full Ni electronic structure
in a basis of maximally localized Wannier functions~\cite{Wannier90:20}
were obtained, followed by an accurate calculation of the spin susceptibilities in TRIQS-TPRF~\cite{TRIQS:15} and subsequent unfolding.
Details are provided in the Supplemental Information (SI)~\cite{SuppMat, GeislerPentcheva-LNOLAO:18, WrobelGeisler:18, GeislerPentcheva-LNOLAO-Resonances:19, MoPa76, MePa89}.

\textit{Ionic relaxation, electronic structure, and interfacial charge transfer. --}
Figure~\ref{fig:OptGeo}(a,b) shows the optimized geometry and layer-resolved density of states of 
epitaxial La$_3$Ni$_2$O$_7$ on LaAlO$_3$(001) and SrTiO$_3$(001).
These prototypical substrates exert
compressive and tensile strain
on the bilayer nickelate (pseudocubic $a_\text{LNO} = 3.83~\AA$), respectively.
The optimized height of the nickelate region,
defined as the distance between the two interfaces,
corresponds to $3/2 \cdot c_\text{LNO}$ of an orthorhombic La$_3$Ni$_2$O$_7$ reference cell,
yielding $c_\text{LNO}=20.13~\AA$ on STO and $20.95~\AA$ on LAO.
Compared to the bulk value of $20.6~\AA$~\cite{Geisler-LNO327-Structure:23}, %
the structure on STO (LAO) is thus considerably contracted (expanded) due to strain.

This trend is also evident in the vertical $A$- and Ni-site distances [Fig.~\ref{fig:OptGeo}(a,b)], %
which can be measured directly in transmission electron microscopy.
The $A$-site distances in the bilayers remain nearly constant at $\sim 3.62~\AA$ on STO,
a notable reduction compared to the bulk value of $3.70~\AA$.
In contrast, these distances vary from $3.66$ to $3.85~\AA$ on LAO.
Across the structural gap between the bilayers,
we find $A$-site distances of $2.86~\AA$ on STO and $2.92~\AA$ on LAO,
compared to the bulk value of $2.90~\AA$.
For the Ni sites,
the bulk separation of $3.94~\AA$
contracts to $3.91~\AA$ on STO
and expands to $3.99~\AA$-$4.04~\AA$ on LAO.
These observations highlight that the interlayer coupling between the Ni sites can be modulated by epitaxial strain.

\begin{figure*}
\begin{center}
\includegraphics[width=\linewidth]{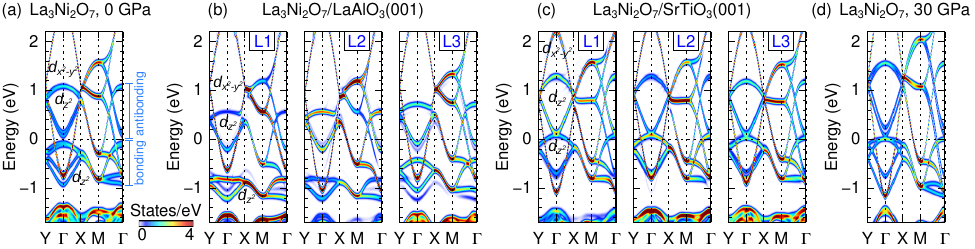}
\caption{\label{fig:Bands}
\textbf{\boldmath Projected band structure of the four considered bilayer nickelate systems.}
Momentum-resolved density of states,
projected on the Ni~$3d$ orbitals in layers L1-L3 %
of (b)~La$_3$Ni$_2$O$_7$/LAO(001) and (c)~La$_3$Ni$_2$O$_7$/STO(001)
(cf.~Fig.~\ref{fig:OptGeo}).
For comparison, consistently obtained bulk Ni~$3d$ bands  %
at (a)~0 and (d)~30~GPa are shown.
Remarkable are the distinct response of the Ni~$e_g$ bands to hydrostatic pressure versus epitaxial strain,
as well as the occupation (depletion) of the (anti)bonding Ni~$3d_{z^2}$ states on LAO (STO)
at the $\Gamma$~point (see also SI).
}
\end{center}
\end{figure*}

The interface geometry considered here [Fig.~\ref{fig:OptGeo}(a,b)]
represents a natural continuation of the two constituent structures
and resembles the layer stacking of related Ruddlesden-Popper cuprate--perovskite nickelate systems
successfully grown in a previous study~\cite{WrobelGeisler:18}.
In [001] direction,
each bilayer of the nickelate consists of three (LaO)$^{1+}$ and two (NiO$_2$)$^{1.5-}$ layers %
yielding a nominal Ni~$3d^{7.5}$ valence.
In contrast,
STO is composed of charge-neutral
(SrO)$^{0}$ and
(TiO$_2$)$^{0}$ layers,
while LAO presents an alternating formal layer polarity of
(LaO)$^{1+}$ and
(AlO$_2$)$^{1-}$.
This raises an intriguing question about
how the different polar discontinuities at the interfaces to the two band insulators are accommodated.

The layer-resolved densities of states in Fig.~\ref{fig:OptGeo}(a,b) reflect the relative band alignment of the constituent oxides.
We find in both systems that
the Fermi energy is clearly located in the substrate band gap. %
Therefore, the Fermi surfaces in Fig.~\ref{fig:OptGeo}(c) are exclusively formed by the Ni~$e_g$ states,
although both substrates exhibit a clear lowering of the conduction-band states towards the interface
[marked by orange and blue arrows in Fig.~\ref{fig:OptGeo}(a,b)].
Specifically, in the STO system, the conduction band edge of the substrate
is identified $\sim 250$~meV above the Fermi level,
and no correlated two-dimensional electron gas (2DEG) forms at the interface.
This constitutes a fundamental difference to infinite-layer nickelate films on STO(001) with ideal interface geometry~\cite{GeislerPentcheva-InfNNO:20, Zhang-IL:20, GeislerPentcheva-NNOCCOSTO:21, Geisler-Rashba-NNOSTOKTO:23}.

In the LAO system,
we find that the polar interface
dopes half an electron per  unit area, %
which is accommodated in La$_3$Ni$_2$O$_7$ %
due to the wide band gap of LAO [Fig.~\ref{fig:OptGeo}(a)].
Hence, the superconductor is charge doped
without the need for chemical doping. %
Tuning the thickness of the nickelate region may be an attractive route to control the density of the doped charge
in the NiO$_2$ planes.
At the interface, %
we observe an expanded $A$-site distance of $3.06~\AA$ relative to the bulk. %
This expansion is attributed to the electrostatic doping
and closely resembles the enhanced La-Sr and Nd-Sr distances across %
$n$-type interfaces in perovskite LaNiO$_3$/STO(001)~\cite{Geisler-LNOSTO:17, ZhangKeimer:14, Hwang:13}
and infinite-layer NdNiO$_2$/STO(001)~\cite{GeislerPentcheva-InfNNO:20}.
Conversely, in the STO system, the corresponding $A$-site distance is
significantly smaller ($2.78~\AA$),
which is a manifestation of the non-polar substrate [Fig.~\ref{fig:OptGeo}(b)]. %

The O-Ni-O bond lengths %
are closely linked to the relative energies and occupation of the Ni~$e_g$ orbitals
and exhibit a significant response to the different substrates (Fig.~\ref{fig:OptGeo}).
On LAO, we identify a general elongation in $z$ direction, %
reaching up to $4.71~\AA$,
while on STO,
they are contracted to as short as $4.01~\AA$,
both relative to the bulk value of $4.27~\AA$.
Conversely, the in-plane O-Ni-O bond lengths amount to $3.81~\AA$ on LAO and $3.93~\AA$ on STO, with a bulk value of $3.85~\AA$.

The overall expanded vertical bond lengths on LAO lower the energy of the Ni~$3d_{z^2}$ orbital relative to Ni~$3d_{x^2-y^2}$,
as evidenced by the projected momentum-resolved density of states,
$A_i(\epsilon, k) = \sum_{n} \, \langle \phi_i \vert \psi_{n,k} \rangle \, \delta(\epsilon_{n,k}-\epsilon)$,
where $\phi_i$ denotes Ni~$3d$ manifolds at different sites~$i$.
We find that this results in the unconventional occupation of the antibonding Ni~$3d_{z^2}$ states around the $\Gamma$ point [Fig~\ref{fig:Bands}(b)], %
which are completely empty in the bulk, even under finite pressure [Fig~\ref{fig:Bands}(a,d)].
This phenomenon is further promoted by the doped charge.
Simultaneously, the bonding Ni~$3d_{z^2}$ states are completely filled
and lowered to $\sim -1$~eV in layer~L1 with considerably reduced bandwidth
[Figs.~\ref{fig:OptGeo}(a), \ref{fig:Bands}(b)],
which coincides with a strong elongation of the corresponding NiO$_6$ octahedra.

Intriguingly, the opposite trend is observed for the STO system,
in which the bonding Ni~$3d_{z^2}$ states form flat bands near the Fermi level
and are partially depleted, similar to pressurized bulk La$_3$Ni$_2$O$_7$ [Fig~\ref{fig:Bands}(c)].
No interfacial charge transfer occurs,
and their energies and band widths remain rather constant throughout all layers.

\begin{figure*}
\begin{center}
\includegraphics[width=\linewidth]{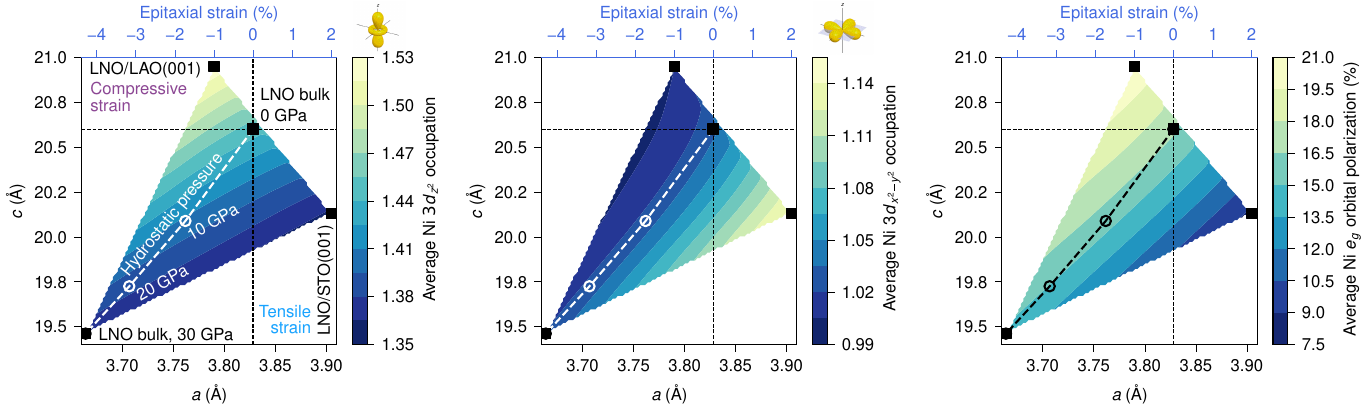}
\caption{\label{fig:Comparison}
\textbf{\boldmath Fundamental difference between hydrostatic pressure and epitaxial strain in bilayer nickelates.}
The panels show the correlations between the optimized lattice parameters %
and the  Ni~$3d$ orbital occupation %
for bulk La$_3$Ni$_2$O$_7$ (LNO) at 0~GPa, 30~GPa, LNO/LAO(001), and LNO/STO(001) (black squares).
For the latter two systems, $a$ is fixed by the substrate, %
which exerts the epitaxial strain $\varepsilon = a/a_\text{LNO} - 1$ on the nickelate.
The dashed line connects the reference points obtained under hydrostatic pressure, %
which we find to explore structural and electronic domains of bilayer nickelates that are orthogonal to those accessible under epitaxial strain.
The colors represent (from left to right)
the average occupation of Ni~$3d_{z^2}$, Ni~$3d_{x^2-y^2}$,
and the resulting Ni $e_g$ orbital polarization, %
interpolated between the exact data points to emphasize the trends.
This highlights that strain provides a much stronger control over the orbital polarization than pressure. %
}
\end{center}
\end{figure*}

\textit{Fermi surface reconstruction. --}
The contrasting response of the electronic structure to compressive versus tensile strain %
results in fundamentally distinct Fermi surfaces [Fig.~\ref{fig:OptGeo}(c)].
Previous work identified the characteristic Fermi surface of La$_3$Ni$_2$O$_7$ to be composed of two sheets:
One of predominantly Ni~$3d_{x^2-y^2}$ character ($\alpha$),
and one with some admixture of Ni~$3d_{z^2}$ ($\beta$)~\cite{Luo-LNO327:23, Gu-LNO327:23, Yang-LNO327:23, Yang-LNO327-ARPES:23}.
Upon application of hydrostatic pressure,
the Fermi surface undergoes a topological transition,
resulting in the emergence of a hole pocket of strong Ni~$3d_{z^2}$ character ($\gamma$)
\cite{Sun-327-Nickelate-SC:23, Luo-LNO327:23, Gu-LNO327:23, Yang-LNO327:23, Lechermann-LNO327:23, Liu-LNO327-OxVacDestructive:23, ZhangDagotto-LNO327:23}.

In the STO system,
we find significant enhancements of both the Ni~$3d_{z^2}$ contribution to the $\alpha$ sheet, %
as well as the Ni~$3d_{x^2-y^2}$ character of the $\beta$ sheet [Fig.~\ref{fig:OptGeo}(c)].
Most importantly, the metallization of the bonding Ni~$3d_{z^2}$ states
results in a Fermi surface topology comparable to the high-$T_c$ superconductor La$_3$Ni$_2$O$_7$ under pressure,
characterized by the emergence of a $\gamma$ hole pocket
susceptible to strain tuning.

In sharp contrast,
the LAO substrate results in a splitting (broadening) of the $\beta$ sheet due to electrostatic doping
and interfacial symmetry breaking,
accompanied by an enhanced Ni~$3d_{z^2}$ contribution.
The $\alpha$ sheet appears considerably more rectangular than in the bulk and displays an enhanced Ni~$3d_{x^2-y^2}$ character.
Additional sheets can be identified around the $\Gamma$ point, %
which stem from the antibonding Ni~$3d_{z^2}$ states,
distinct from the STO case.

These findings establish that strain and interface polarity provide substantial control over the Fermi surface topology,
extending far beyond pressure effects. %
Notably, the Fermi surface reconstruction is not accompanied by a suppression of the NiO$_6$ octahedral rotations [Fig.~\ref{fig:OptGeo}(a,b)],
not even under compressive strain,
which marks an important difference to pressurized bulk La$_3$Ni$_2$O$_7$~\cite{Sun-327-Nickelate-SC:23, Geisler-LNO327-Structure:23, ZhangDagotto-LNO327:23, Wang-LNO327-I4mmm:23}. %
This opportunity to tune the electronic structure
independently of the octahedral degrees of freedom
grants access to a more profound understanding of superconductivity in bilayer nickelates.

\textit{Fundamental differences between hydrostatic pressure and epitaxial strain. --}
The essence of these results is compiled in Fig.~\ref{fig:Comparison}. %
The panels show the correlations between the optimized lattice parameters %
and the average occupation of the Ni~$3d$ orbitals
for bulk La$_3$Ni$_2$O$_7$ at 0~GPa, 30~GPa, La$_3$Ni$_2$O$_7$/LAO(001), and La$_3$Ni$_2$O$_7$/STO(001).
In the case of orthorhombic bulk La$_3$Ni$_2$O$_7$, we show the pseudocubic $a$ value.
For the heterostructures, $a$ is fixed by the substrate,
and we show the optimized $c_\text{LNO}$ value  described above. %

The figure establishes
that epitaxial strain explores structural and electronic domains
of the bilayer nickelates that are orthogonal to those accessible under external pressure.
Hydrostatic pressure results in the simultaneous compression of  $a$ and $c$, %
corresponding to the lower-left quadrant in each panel of Fig.~\ref{fig:Comparison}.
This reduces the Ni~$3d_{z^2}$ occupation from $1.45$ (0~GPa) to $1.37$ (30~GPa),
while maintaining an approximately constant Ni~$3d_{x^2-y^2}$ occupation ($\sim 1.03$).
In sharp contrast, epitaxial strain leads to an opposite evolution of $a$ and $c$,
represented by the upper-left and lower-right quadrants in Fig.~\ref{fig:Comparison},
which drives a significant charge transfer from Ni~$3d_{z^2}$ to Ni~$3d_{x^2-y^2}$ for tensile strain, and vice versa for compressive strain.
This allows to control the Ni~$e_g$ orbital polarization
$(3d_{z^2}-3d_{x^2-y^2})/(3d_{z^2}+3d_{x^2-y^2})$~\cite{GeislerPentcheva-LNOLAO-Resonances:19}
over a considerably larger interval than hydrostatic pressure.
Here, we find it to range from $\sim 8\%$ on STO to $\sim 21\%$ on LAO (Fig.~\ref{fig:Comparison}).

Simultaneously, Fig.~\ref{fig:Comparison} compares the characteristic length scales of the different approaches. %
The values for $a$ ($c$) obtained due to epitaxial strain on LAO (STO) %
require $\sim10$~GPa external pressure.
Conversely, application of $30$~GPa results in a basal contraction %
corresponding to $\sim 4\%$ compressive strain,
which is readily achievable in experiment.

\begin{figure}[b]
\begin{center}
\includegraphics[width=\linewidth]{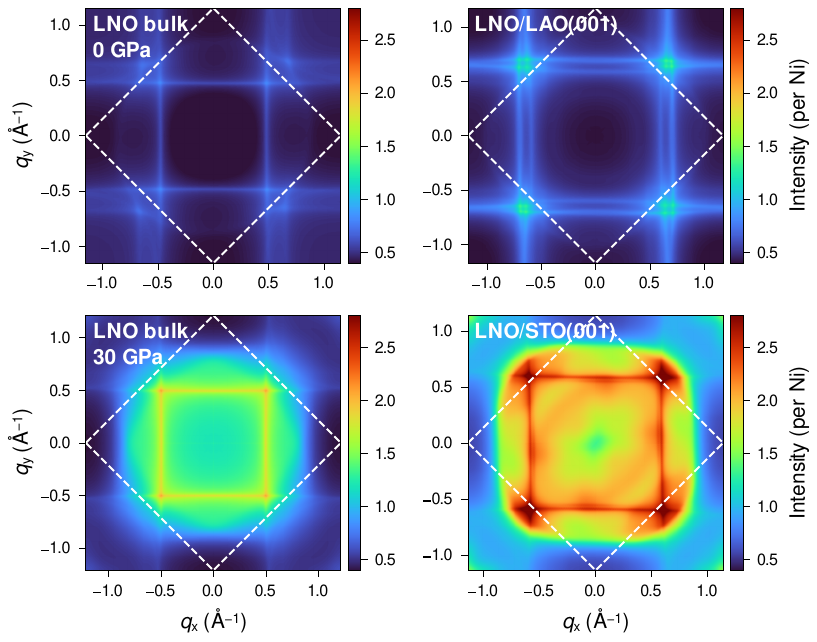}
\caption{\label{fig:Chi}
\textbf{\boldmath Spin-spin correlation functions $\chi^\text{RPA}(\mathbf{q})$.}
The results have been unfolded from the $\sqrt{2}\times\sqrt{2}$ (dashed lines) to the full Brillouin zone. %
Strong enhancements particularly of the peaks along the $(\pi,\pi)$ direction can be observed in the strained systems
that exceed the effect of external pressure considerably for LNO/STO(001).
\vspace{-1.25em}
}
\end{center}
\end{figure}

\textit{Spin susceptibility enhancement. --}
The superconducting pairing is determined by the dynamical spin susceptibility $\chi^\text{RPA}(\mathbf{q})$, %
which we compare in Fig.~\ref{fig:Chi} %
considering a Kanamori-type interaction vertex ($U=0.8$, $J=0.4$~eV) in the random phase approximation (RPA).
The susceptibility of bulk LNO at ambient pressure is dominated by intraorbital $3d_{x^2-y^2}$ scattering. %
We observe qualitative and quantitative changes due to strain,
specifically considerably enhanced intraorbital $3d_{z^2}$ and interorbital $3d_{z^2}$-$3d_{x^2-y^2}$ contributions that clearly surpass the $3d_{x^2-y^2}$ channel (Fig.~\ref{fig:Chi}; see SI for the orbital-resolved susceptibilities).
On LAO, favorable Fermi surface nesting with the antibonding $3d_{z^2}$ states results in a $1.6$-fold increase of the susceptibility peaks along the $(\pi,\pi)$ direction
relative to ambient-pressure bulk LNO.
In pressurized bulk LNO, the enhancement is more pronounced ($\times 2.5$),
reflecting strong intraorbital $3d_{z^2}$ scattering from $\gamma$ pocket nesting~\cite{Luo-LNO327:23}.
Interestingly, we identify the strongest amplification on STO ($\times 4$)
due to the coupling of the $\alpha$ and $\beta$ sheets with the emerging $\gamma$ pocket,
which suggests the potential for an even higher $T_c$.

\textit{Summary. --}
We investigated %
the impact of epitaxial strain and interface polarity %
in La$_3$Ni$_2$O$_{7}$ on LaAlO$_3$(001) and SrTiO$_3$(001) %
by performing first-principles simulations including a Coulomb repulsion term.
For La$_3$Ni$_2$O$_{7}$/LaAlO$_3$(001), compressive strain in conjunction with electron doping across the explicitly treated interface
results in the unconventional occupation of the antibonding Ni~$3d_{z^2}$ states.
In sharp contrast, no charge transfer was observed for La$_3$Ni$_2$O$_{7}$/SrTiO$_3$(001).
Intriguingly, tensile strain was found to drive a metallization of the bonding Ni $3d_{z^2}$ states %
that yields a Fermi surface topology resembling bulk La$_3$Ni$_2$O$_{7}$ under high pressure,
yet with a strongly enhanced spin susceptibility.
We demonstrated that such transitions in the electronic structure, considered to be key for high-$T_c$ superconductivity in bilayer nickelates,
are decoupled from quenching the octahedral rotations.
In addition, we identified the fundamental differences between hydrostatic pressure and epitaxial strain
regarding the crystal geometry and Ni~$3d$ orbital occupation,
and established that strain provides a much stronger control over the Ni~$e_g$ orbital polarization.

The great promise of high-$T_c$ nickelate superconductivity will only be realized if structures favorable for electron pairing can be created at ambient pressure.
Our results show that epitaxial strain is an excellent avenue to explore in this direction.
Already moderate strain as exerted by the technologically relevant substrates LaAlO$_3$ and SrTiO$_3$
is predicted to alter the electronic properties and spin fluctuations considerably due to the strong response of the Ni orbitals.
We particularly emphasized the similarity of La$_3$Ni$_2$O$_7$ under tensile strain to the high-$T_c$ pressurized bulk compound,
suggesting the potential for superconductivity at ambient pressure.
Finally, the proposed strategy avoids chemical doping and concomitant disorder, and may hence allow even higher critical temperatures than heretofore achieved.

\textit{Notes added. --}
After submission, ambient-pressure superconductivity has been reported in capped bilayer nickelate thin films on SrLaAlO$_4$(001)~\cite{Ko-LNO-ComprStrain-SC:24, Liu-LNO-CompressiveStrain-SC:25, Zhou-LNO-Strain:25}.
Earlier efforts did not find a superconducting transition~\cite{Cui-LNO-Strain:24}. %
Our results suggest the corresponding superconductivity mechanism to be fundamentally distinct from La$_3$Ni$_2$O$_7$ under high pressure due to interfacial charge transfer and the involvement of the \textit{antibonding} Ni states;
the observed reduced $T_c \sim 40$~K agrees with the trends established by our susceptibility calculations.
The concomitant presence of octahedral rotations has been confirmed by high-resolution transmission electron microscopy~\cite{Bhatt-LNO-Strain:25}
and independent theoretical work for the strained bulk compound~\cite{BhattaJia-LNO-Strain:25},
as has the identified Fermi surface reconstruction~\cite{Zhao-LNO-Strain:24}.

\begin{acknowledgments}
\textit{Acknowledgments. --} This work was supported by the National Science Foundation, Grant No.~NSF-DMR-2118718.
\end{acknowledgments}

\end{document}